\documentclass[useAMS,usenatbib]{mn2e}
\usepackage{times}
\usepackage{graphics,epsfig}
\usepackage{graphicx}
\usepackage{amssymb}
\usepackage{txfonts}

\newcommand{\kms}{km\,s$^{-1}$} 
\newcommand{\ts}  {\ensuremath{{\rm T}_{\rm s}}}
\newcommand{\tnhi}  {\ensuremath{10^{20}}}
\newcommand{\nhi}{\ensuremath{N}_{\rm HI}}

\newcommand{\cm}{cm$^{-2}$}
\newcommand{\hi}{H\,{\sc i}}
\newcommand{\hii}{H\,{\sc i} 21\,cm}

\title[A new low ${\rm T_s}$ DLA at high redshift]{A search for H{\sc i}\,21cm absorption 
towards a radio-selected quasar sample II: a new low spin temperature DLA at high redshift}
\author[N. Kanekar et al.]{Nissim Kanekar$^{1}$\thanks{E-mail: nkanekar@ncra.tifr.res.in~(NK)},
Sara L. Ellison$^2$, Emmanuel Momjian$^{3}$, Brian A. York$^{4}$, and Max Pettini$^{5,6}$\\
$^{1}$National Centre for Radio Astrophysics, TIFR, Post Bag 3, Ganeshkhind, Pune 411 007, India;\\
$^{2}$Department of Physics and Astronomy, University of Victoria, B.C., V8P 1A1, Canada; \\
$^{3}$National Radio Astronomy Observatory, 1003 Lopezville Road, Socorro, NM 87801, USA; \\
$^{4}$Space Telescope Science Institute, 3700 San Martin Drive, Baltimore, MD 21218, USA; \\
$^{5}$Institute of Astronomy, Madingley Rd., Cambridge CB3 0HA, UK; \\
$^{6}$ International Centre for Radio Astronomy Research, University of Western Australia, 
7 Fairway, Crawley WA 6009, Australia}

\begin{document}
\date{Accepted yyyy month dd. Received yyyy month dd; in original form yyyy month dd}


\maketitle

\label{firstpage}

\begin{abstract}

We report results from a deep search for redshifted H{\sc i}\,21cm absorption from 
eight damped Lyman-$\alpha$ absorbers (DLAs) detected in our earlier optical survey 
for DLAs towards radio-loud quasars.  H{\sc i}\,21cm absorption was detected from 
the $z = 2.192$ DLA towards TXS2039+187, only the sixth case of such a detection at 
$z > 2$, while upper limits on the H{\sc i}\,21cm optical depth were obtained in six 
other DLAs at $z > 2$. Our detection of H{\sc i}\,21cm absorption in the eighth system, 
at $z = 2.289$ towards TXS0311+430, has been reported earlier. We also present high 
spatial resolution images of the background quasars at frequencies close to the 
redshifted H{\sc i}\,21cm line frequency, allowing us to estimate the covering factor 
of each DLA, and then determine its spin temperature ${\rm T_s}$. For three non-detections 
of H{\sc i}\,21cm absorption, we obtain strong lower limits on the spin temperature, 
${\rm T_s} \gtrsim 790$~K, similar to the bulk of the high-$z$ DLA population; three other 
DLAs yield weak lower limits, $\ts > 140-400$~K. However, for the two DLAs with 
detections of H{\sc i}\,21cm absorption, the derived spin temperatures are both low
${\rm T_s} = (160 \pm 35) \times (f/0.35)$~K for the $z = 2.192$ DLA towards TXS2039+187 
and ${\rm T_s} = (72 \pm 18) \times (f/0.52)$~K for the $z = 2.289$ DLA towards TXS0311+430.
These are the first two DLAs at $z > 1$ with ${\rm T_s}$ values comparable to those obtained 
in local spiral galaxies. Based on the observed correlations between spin temperature 
and metallicity and velocity spread and metallicity in DLAs, we suggest that the hosts 
of the two absorbers are likely to be massive, high-metallicity galaxies.

\end{abstract}


\section{Introduction}
\label{sec:intro}

Selected through quasar absorption surveys, and with neutral hydrogen (\hi) column 
densities $\geq 2 \times 10^{20}$~\cm, damped Lyman-$\alpha$ absorbers (DLAs) are 
the high-redshift analogues of gas-rich galaxies in the local Universe. The nature of 
high-$z$ DLAs and their redshift evolution have long been subjects of much interest 
\citep[e.g.][]{wolfe05}. Unfortunately, our information about these systems is 
mostly based on absorption studies, as it has proved very difficult to detect the absorber 
hosts in the presence of the bright background quasar 
\citep[although see][]{fumagalli10,peroux12,noterdaeme12,krogager12}. 
Absorption studies only sample a narrow pencil beam through the intervening galaxy and 
hence often lend themselves to ambiguous interpretation. Further, the strong low-ionization 
metal lines commonly detected in DLAs rarely provide direct information about 
local physical conditions (e.g.  density, temperature, pressure) in the interstellar medium 
(ISM) of the absorbers. Such information is only available for a few DLAs, systems with 
simple velocity structure \citep[e.g.][]{pettini01,carswell12}, molecular hydrogen (H$_2$) absorption
\citep[e.g.][]{ledoux03,srianand05}, neutral carbon absorption 
\citep[e.g.][]{jorgenson09,carswell11}, Si{\sc ii}* absorption \citep{howk05}, etc. As a 
result, despite years of detailed studies, our understanding of the host galaxies of 
high-$z$ DLAs remains quite limited.


For DLAs occulting radio-loud quasars, a comparison between the \hi\ column density 
measured from the Lyman-$\alpha$ absorption profile and the \hii\ equivalent width 
can be used to obtain the spin temperature ($\ts$) of the neutral gas \citep[e.g.][]{kanekar04}.
This provides information about the distribution of \hi\ in different phases of 
the neutral ISM, and is thus one of the few direct tracers of local conditions in DLAs.
Of course, such studies require the assumption that the \hi\ column density towards the 
optical quasar is the same as that towards the radio source. While the latter is usually 
far more extended than the optical quasar, especially at low frequencies, $\lesssim 1$~GHz, 
very long baseline interferometry (VLBI) studies at frequencies similar to the redshifted \hii\ 
line frequency allow one to estimate the fraction of radio flux density arising from the 
compact radio core, and thus the low-frequency covering factor $f$ of the DLA 
\citep[e.g.][]{wolfe76,briggs83,kanekar09a}. Estimating the DLA spin temperature thus 
requires three measurements: (1)~the \hi\ column density from the Lyman-$\alpha$ line,
(2)~the \hii\ optical depth, and (3)~the covering factor $f$ from low-frequency VLBI
studies (which also requires unambiguous identification of the radio quasar core). Note that 
this is still based on the assumption that the \hi\ column density 
measured towards the optical quasar is the same as towards the radio quasar core \citep[i.e. 
over a spatial extent of a few hundred parsec at the absorber redshift][]{kanekar09a}. 
Comparisons between \hi\ column densities measured from the Lyman-$\alpha$ profile and 
from \hii\ emission studies along similar sightlines in the Galaxy have found very 
similar values, with a mean ratio of unity and a dispersion of $\approx 10$\% \citep{wakker11}.
A similar comparison in the $z \approx 0.009$ DLA towards SBS~1549+593 also yielded 
good agreement between the \hi\ columns estimated from the two very different methods 
\citep{bowen01a,chengalur02}.

Unfortunately, the number of high-$z$ absorbers with spin temperature estimates is still 
quite small. While more than a thousand DLAs are now known at $z > 2$ 
\citep[e.g.][]{prochaska05,prochaska09,noterdaeme09}, there were, prior to this survey,
only four DLAs with detections of \hii\ absorption at these redshifts 
\citep{wolfe85,kanekar06,kanekar07,srianand10}, besides about twenty non-detections 
of \hii\ absorption giving strong lower limits on the spin temperature 
\citep[e.g.][]{carilli96,kanekar03,kanekar09c,srianand12}. The main reason for the lack of 
$\ts$ estimates in high-$z$ DLAs is that very few DLAs are known towards bright 
radio-loud quasars. To address this issue, we have carried out an optical survey 
of 45 quasars selected on the basis of their low-frequency flux density; 
this resulted in the detection of eight new DLAs at $z \gtrsim 2$ \citep{ellison08}.
Our first detection of \hii\ absorption from this sample was reported in \citet{york07};
here, we present results from a search for redshifted \hii\ absorption in the full
sample, which has yielded the first two low spin temperature DLAs at high redshifts.

\begin{figure*}
\centering
\epsfig{file=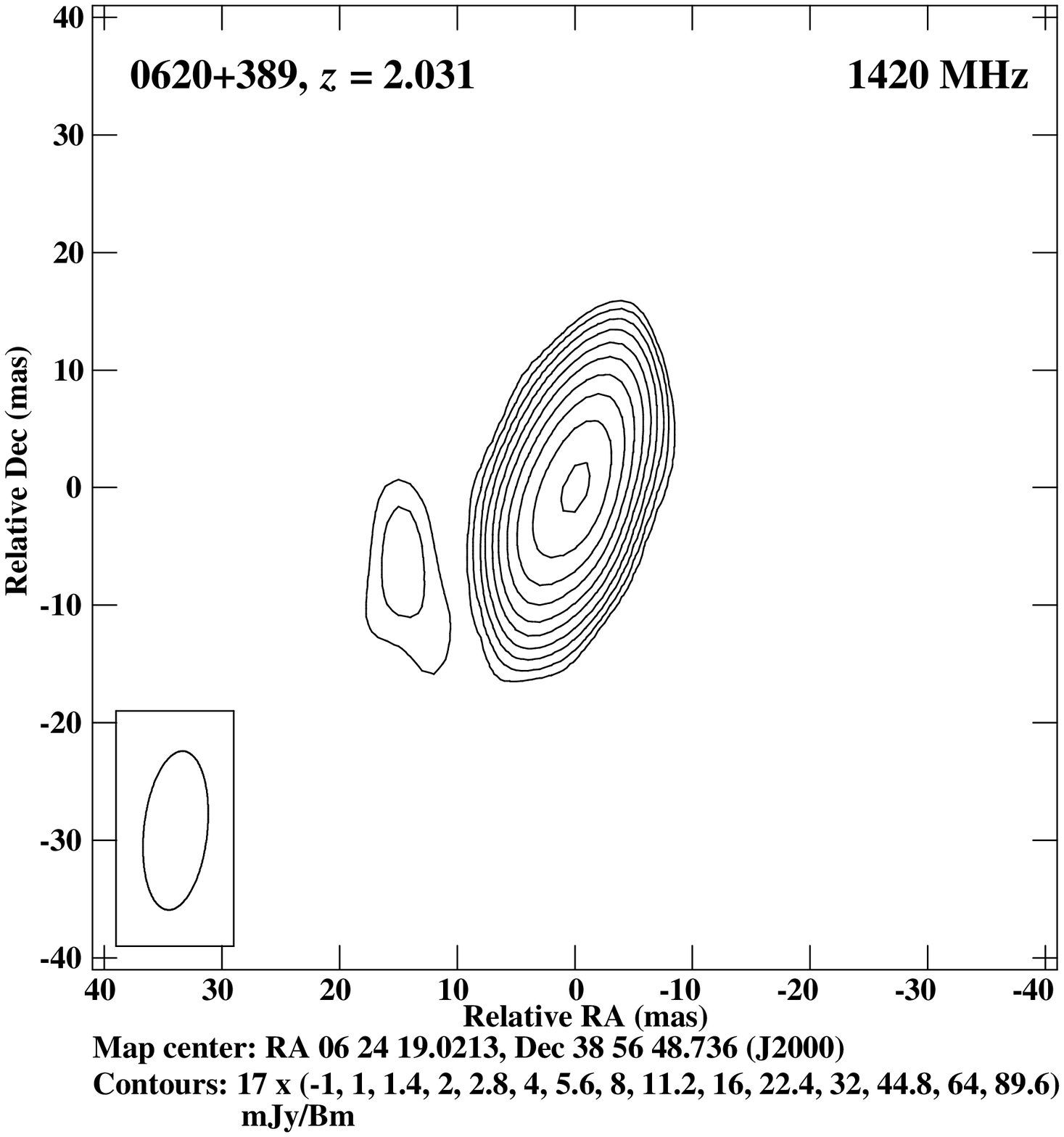,height=2.2truein,width=2.2truein}
\epsfig{file=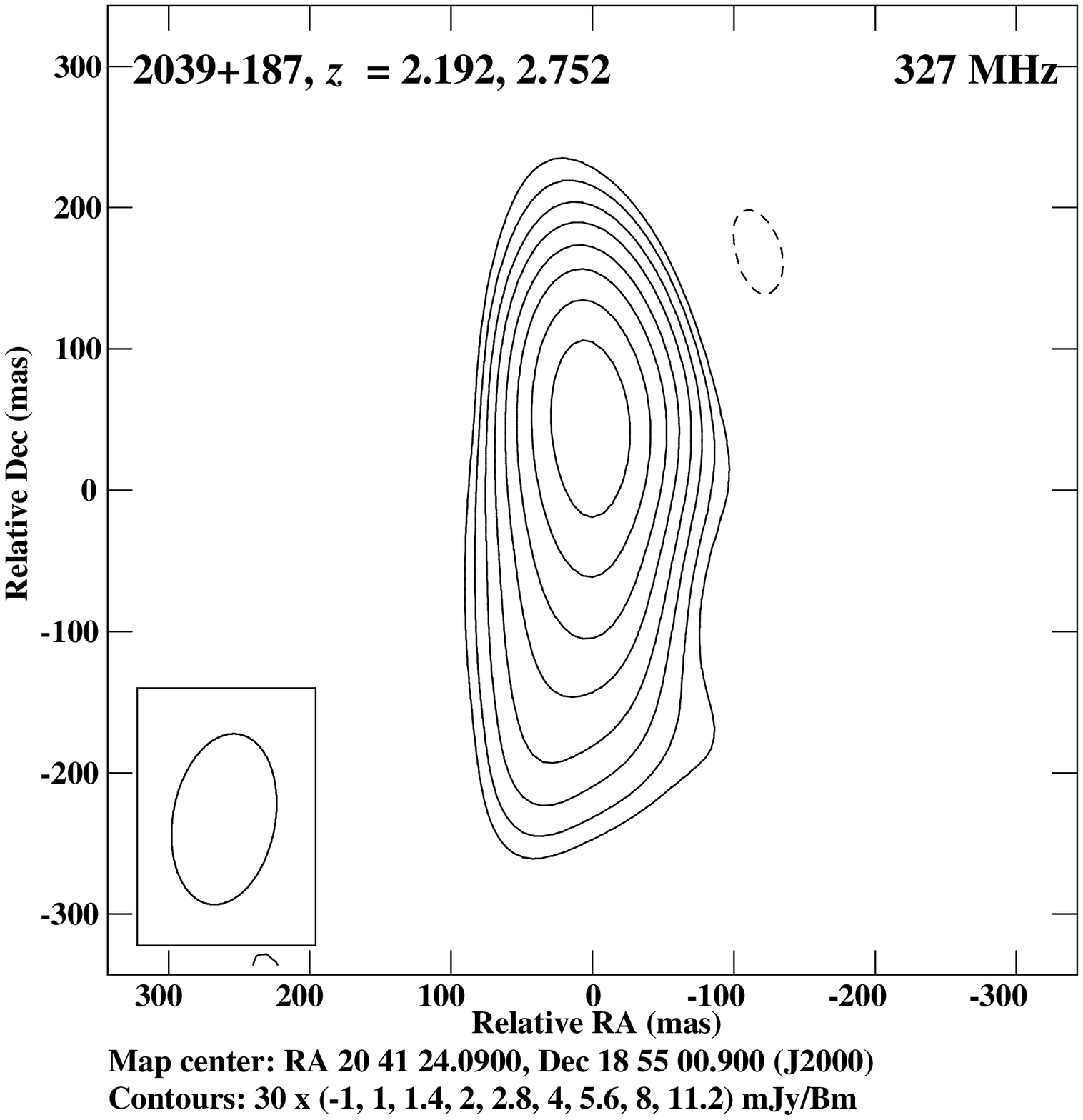,height=2.2truein,width=2.2truein}
\epsfig{file=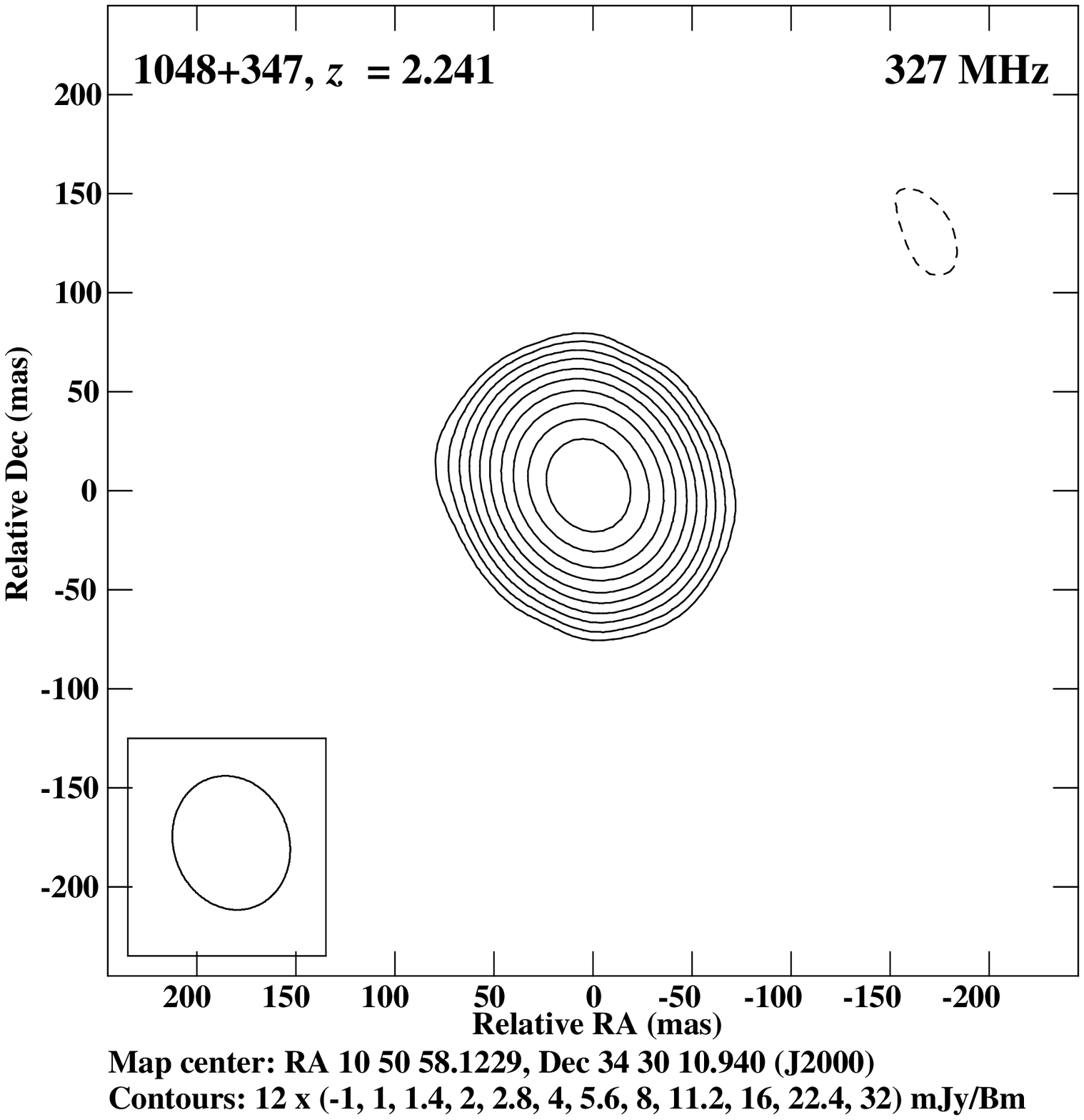,height=2.2truein,width=2.2truein}
\epsfig{file=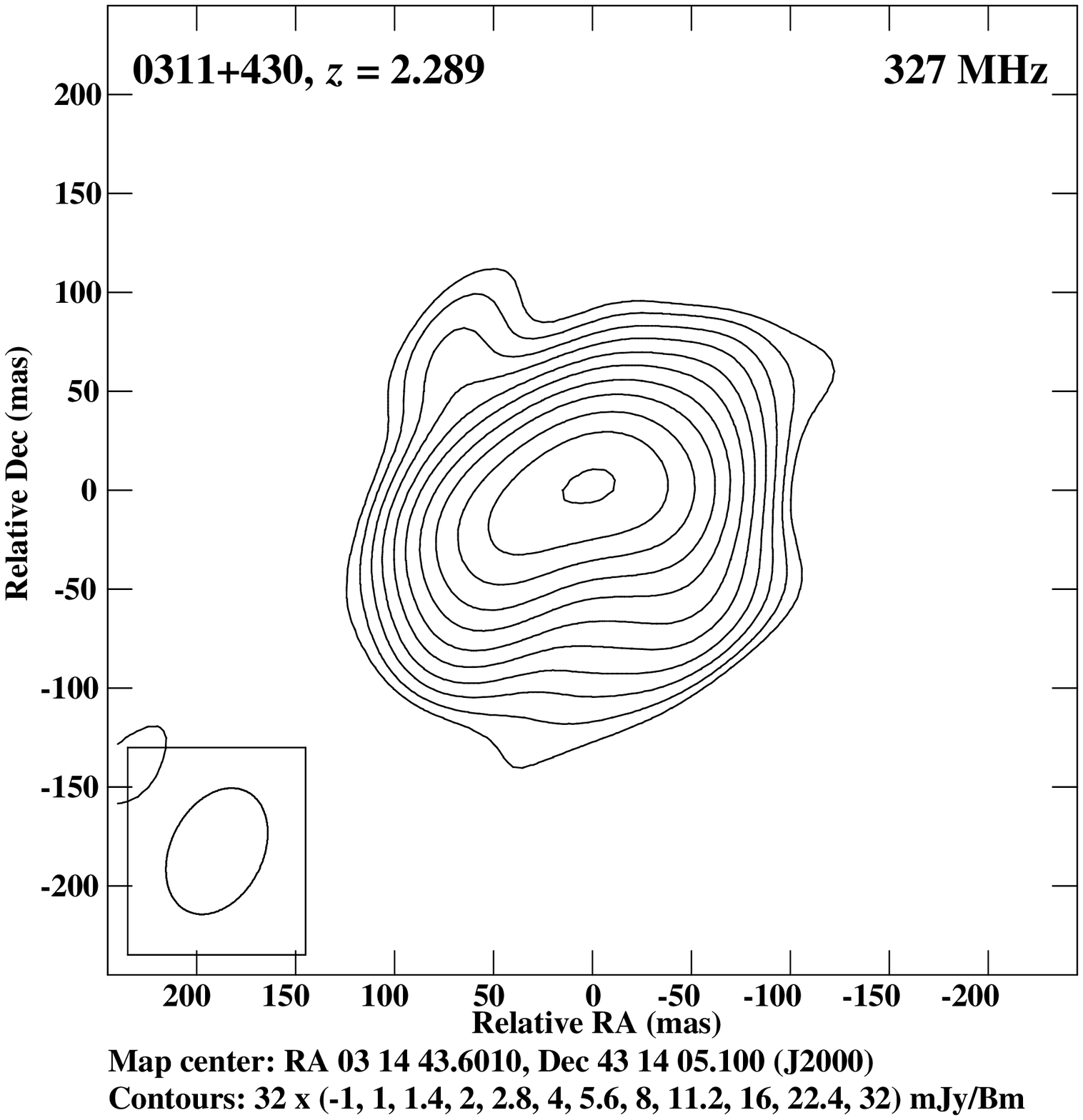,height=2.2truein,width=2.2truein}
\epsfig{file=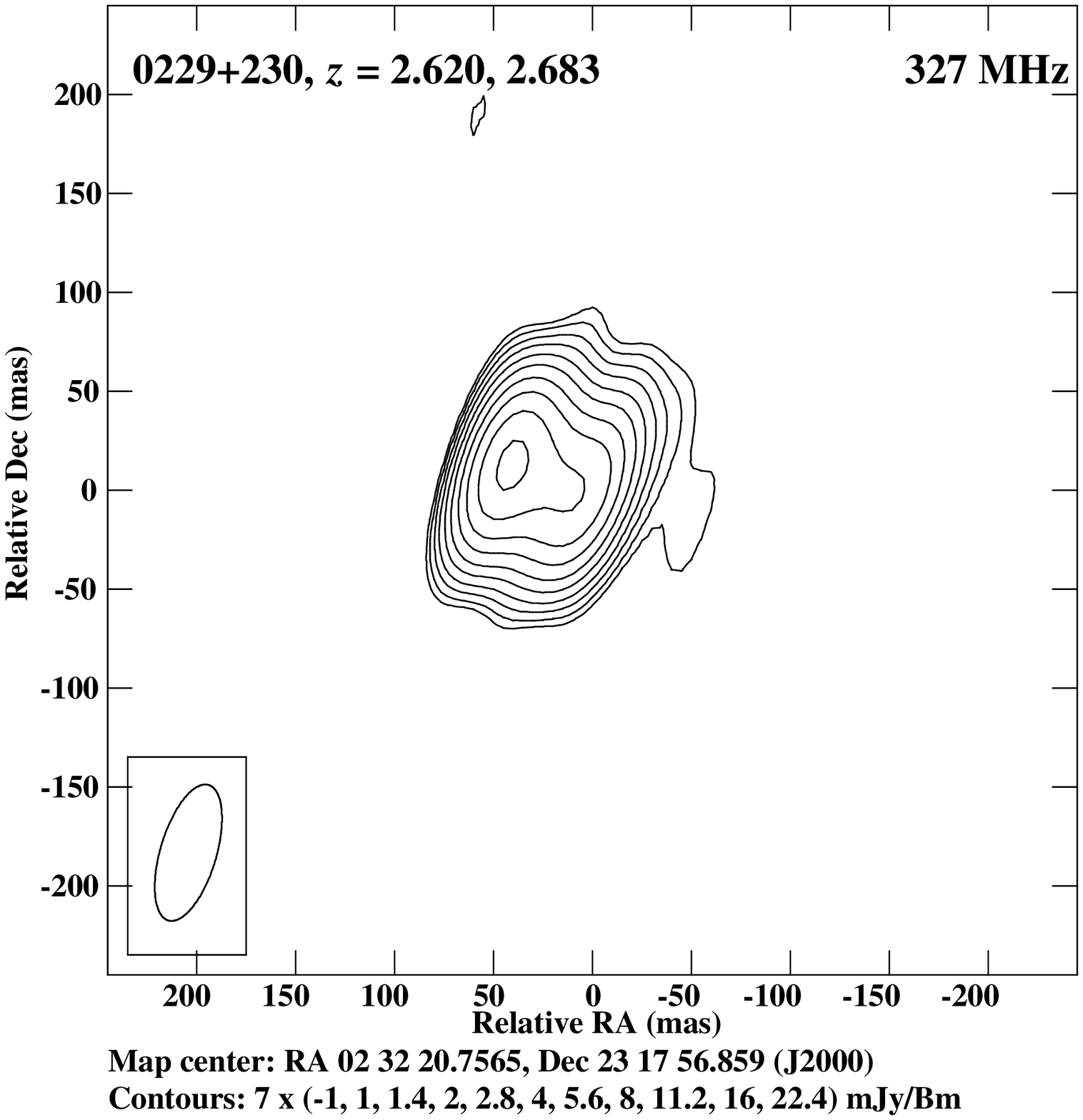,height=2.2truein,width=2.2truein}
\epsfig{file=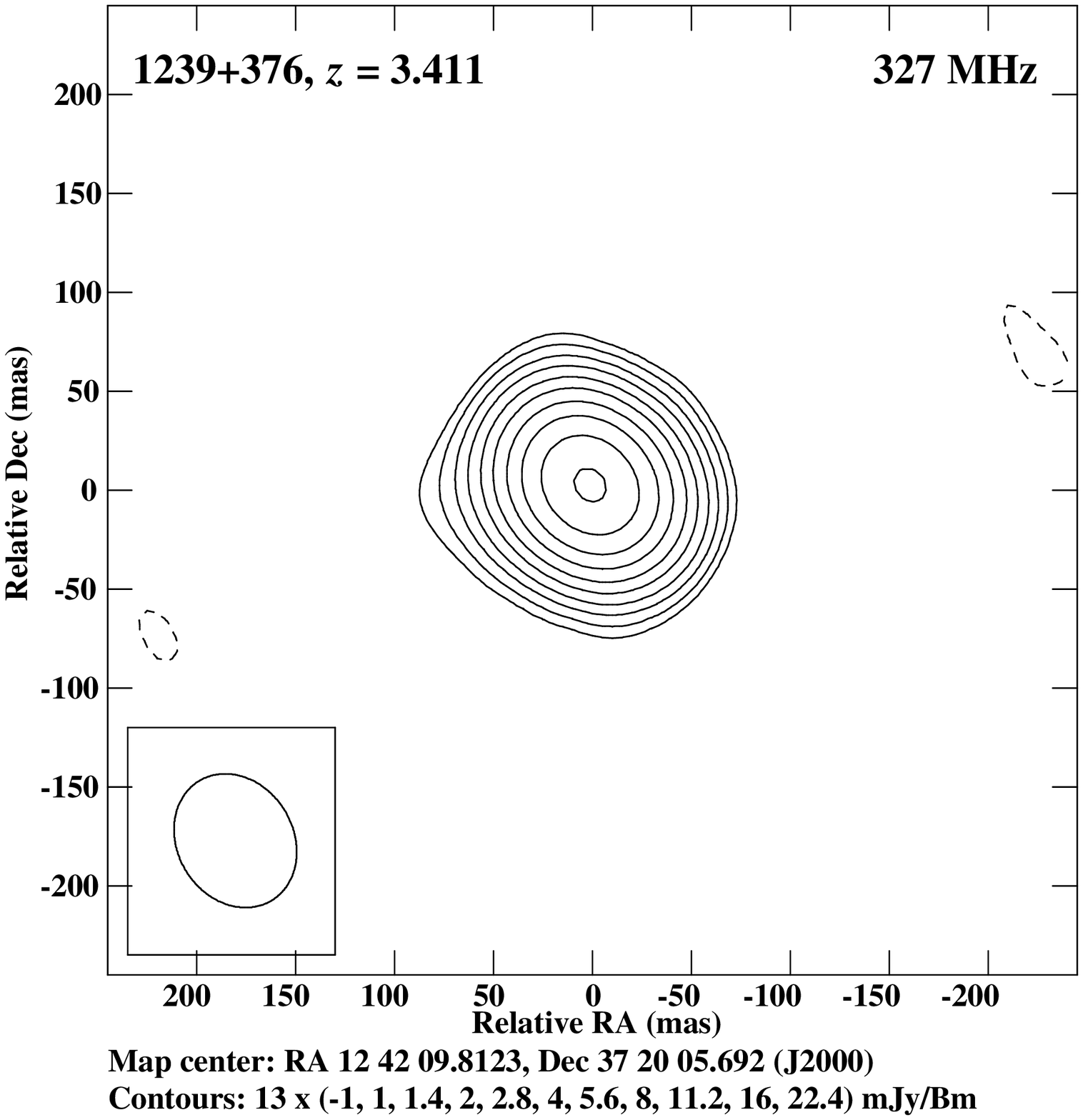,height=2.2truein,width=2.2truein}
\caption{VLBA images of the compact radio structure of the six background quasars; note that 
two of the quasar sightlines contain two DLAs. The quasar name, DLA redshift and observing 
frequency are indicated at the top of each panel. 
}
\label{fig:vlba}
\end{figure*}

\setcounter{table}{0}

\begin{table*}
\caption{Results from VLBA low-frequency imaging of quasars behind the DLAs of our sample.
}
\begin{center}
\begin{tabular}{|c|c|c|c|c|c|c|c|c|c|c|c|}
\hline
TXS  & $z_{\rm DLA}$ & $\nu_{\rm 21cm}$ & $\nu_{\rm VLBA}$ & Beam             & RMS             & S$_{\rm VLBA}$ & $S_{\rm int}$ & Angular size      & Spatial extent  & $f$\\
name &               &     MHz          & MHz              & mas $\times$ mas & mJy bm$^{-1}$ & Jy             &     Jy        &  mas $\times$ mas & pc $\times$ pc    & \\
\hline
	 & && & & & & & & & \\
0620+389 & 2.031 & 468.63 & 1400 & $13.6 \times 5.4$ & 5.1 & 0.64 & 0.81 & $9.7^{+0.3}_{-0.4} \times 0.7^{+1.2}_{-0.7}$ & ~~$82^{+3}_{-3} \times 6^{+10}_{-6}$~~ & 0.79 \\
	 & && & & & & & & & \\
2039+187 & 2.192 & 444.99 & 327~ & ~$122 \times 72$~ & 11~ & 0.68 & 1.92 & $129^{+7}_{-8} \times 34^{+8}_{-12}$~ & $1081^{+59}_{-67} \times 285^{+67}_{-101}$ & 0.35$^a$ \\
	 & ~~~~~ & ~~~~~~ & ~~~~ & ~~~~~~~~~~~~~~~~~ & ~~~ & 0.39 & ~~~~ & $190^{+24}_{-25} \times 85^{+14}_{-16}$~ & ~~~~~~~~~~~~~~~~~	& ~~~~ \\
	 & && & & & & & & & \\
1048+347 & 2.241 & 438.26 & 327~ & ~~$69 \times 58$~ & 2.7 & 0.40 & 0.58 & ~$23^{+3}_{-3} \times 8.6^{+5.1}_{-8.6}$ & ~$192^{+25}_{-25} \times 72^{+43}_{-72}$~ & 0.69 \\
	 & && & & & & & & & \\
0311+430 & 2.289 & 431.87 & 327~ & ~~$67 \times 47$~ & 14~ & 3.12 & 5.96 & $106^{+2}_{-2} \times 55^{+2}_{-2}$~ & ~$883^{+17}_{-17} \times 458^{+17}_{-17}$ & 0.52$^a$ \\
	 & ~~~~~ & ~~~~~~ & ~~~~ & ~~~~~~~~~~~~~~~~~ & ~~~ & 1.21 & ~~~~ & $100^{+7}_{-7} \times 86^{+7}_{-6}$~ & ~~~~~~~~~~~~~~~~~	& ~~~~ \\
	 & && & & & & & & & \\
0229+230 & 2.620 & 392.38 & 327~ & ~~$71 \times 28$~ & 1.7 & 0.18 & 0.59 & ~$24^{+2}_{-24} \times 0.0^{+24}_{-0}$ & ~$195^{+16}_{-195} \times 0^{+195}_{-0}$~~ & 0.30$^a$ \\
	 & ~~~~~ & ~~~~~~ & ~~~~ & ~~~~~~~~~~~~~~~~~ & ~~~ & 0.19 & ~~~~ & ~$41^{+2}_{-3} \times 18^{+5}_{-7}$~ & ~~~~~~~~~~~~~~~~~	& ~~~~ \\
	 & && & & & & & & & \\
0229+230 & 2.683 & 385.67 & 327~ & ~~$71 \times 28$~ & 1.7 & 0.18 & 0.59 & ~$24^{+2}_{-24} \times 0.0^{+24}_{-0}$ & ~$194^{+16}_{-194} \times 0^{+194}_{-0}$~~ & 0.30$^a$ \\
	 & ~~~~~ & ~~~~~~ & ~~~~ & ~~~~~~~~~~~~~~~~~ & ~~~ & 0.19 & ~~~~ & ~$41^{+2}_{-3} \times 18^{+5}_{-7}$~ & ~~~~~~~~~~~~~~~~~	& ~~~~ \\
	 & && & & & & & & & \\
2039+187 & 2.752 & 378.58 & 327~ & ~$122 \times 72$~ & 11~ & 0.68 & 1.92 & $129^{+7}_{-8} \times 34^{+8}_{-12}$~ & $1034^{+56}_{-64} \times 272^{+64}_{-96}$ & 0.35$^a$ \\
	 & ~~~~~ & ~~~~~~ & ~~~~ & ~~~~~~~~~~~~~~~~~ & ~~~ & 0.39 & ~~~~ & $190^{+24}_{-25} \times 85^{+14}_{-16}$~ & ~~~~~~~~~~~~~~~~~ & ~~~~ \\
	 & && & & & & & & & \\
1239+376 & 3.411 & 322.01 & 327~ & ~~$71 \times 59$~ & 3.8 & 0.34 & 0.62 & ~$30^{+3}_{-5} \times 12^{+7}_{-12}$~ & ~$226^{+23}_{-38} \times 90^{+53}_{-90}$ & 0.54 \\
	 & && & & & & & & & \\
\hline
\end{tabular}
\label{table:vlba}
\end{center}
Notes: $^a$~Note that the more compact component was identified as the radio core in all five cases where 
a 2-Gaussian model was fit to the VLBA image. The covering factors for the DLAs towards TXS0229+230 and TXS2039+187 
assume that only the radio core is covered. If the second VLBA component is also covered by the foreground 
absorber, the covering factors would be $f=0.56$ (TXS2039+187), $f=0.63$ (TXS0229+230) and $f=0.73$ (TXS0311+430). 
In the case of TXS0311+430, the second VLBA component is located $1.42''$ to the south-west, corresponding to 
a distance of $\approx 11.7$~kpc away at the DLA redshift. This is unlikely to be covered by the foreground galaxy, 
unless it is a large disk.
\end{table*}

\section{The Observations}
\label{sec:sample}

Our search for redshifted \hii\ absorption in the eight DLAs of our sample (with two 
sightlines, towards TXS0229+230 and TXS2039+187, containing two DLAs; see 
Table~\ref{table:vlba}) was carried out with the Green Bank Telescope (GBT; six DLAs), 
the Giant Metrewave Radio Telescope (GMRT; 1~DLA) and the Westerbork Synthesis Radio 
Telescope (WSRT; 1~DLA). The GBT observations (proposals AGBT06B-042, AGBT07B-008, 
AGBT08A-076 and AGBT11B-221) were carried out between 2007 and 2011, using the 
PF1-450~MHz and PF1-342~MHz receivers, 2~linear polarizations and the Spectral Processor. 
A 2.5~MHz band, sub-divided into 1024~channels, was used for all sources except TXS0229+230,
whose two DLAs were observed simultaneously, using two 2.5~MHz bands, each sub-divided 
into 512~channels. This yielded a velocity resolution of $\sim 3-8$~\kms, after Hanning 
smoothing and resampling. 
Standard position-switching, with On and Off~times of 5~minutes each, was used for calibration, 
with system temperatures measured using a blinking noise diode. The two DLAs (towards TXS0311+430 
and TXS2039+187) with absorption features near the expected line frequency were re-observed 
on multiple occasions with a bandwidth of 1.25~MHz. In both cases, the reality of the \hii\ 
absorption was confirmed by the detection of the doppler shift due to the Earth's motion
between the different runs.

The GMRT observations of TXS1239+376 were carried out in February~2011; earlier runs in 2007 
and 2009 were found to be affected by weak radio frequency interference (RFI) and will not be 
discussed further. A total of 28 antennas were available for the observations, which used the 
GMRT Software Backend and a bandwidth of 2.08~MHz, sub-divided into 512~channels; this gave 
a velocity resolution of 7.6~\kms, after Hanning smoothing and resampling. The calibrators 
3C147 and 3C286 were used to calibrate the flux density scale, the system gain and the 
system bandpass. 

Finally, the $z = 2.752$ DLA toward TXS2039+187 was observed with the WSRT, in September~2008, 
with a bandwidth of 2.5~MHz and 2048 channels (i.e. a velocity resolution of 1.9~\kms\ after 
Hanning smoothing and resampling).  The flux density scale and system bandpass were 
calibrated with observations of 3C286 and 3C48.

We also used the Very Long Baseline Array (VLBA) to obtain high-resolution images of the 
radio continuum emission of the six background quasars of our sample, to estimate the 
absorber covering factor $f$ at a frequency close to that of the redshifted \hii\ line. 
Five quasars were observed with the VLBA 327~MHz receivers in 2008 and 2009 (proposal 
BK153), using a bandwidth 12~MHz, sub-divided into 32 spectral channels and with 2 polarizations, 
2-bit sampling and on-source times of $\approx 2$~hours. A strong fringe finder 
(3C454.3, 3C84, 3C147, 3C286 or 3C345) was observed during each run for bandpass calibration; 
phase referencing was not used. For TXS0620+389, we reduced an archival VLBA 1420~MHz 
dataset from 2002; this had a bandwidth of 8~MHz, sub-divided into 16 channels, with 
2 polarizations and an on-source time of 8 minutes.

\begin{table*}
\begin{center}
\caption{Results of the \hii\ absorption spectroscopy and estimates of the spin temperature
\label{table:sample}}
\begin{tabular}{|c|c|c|c|c|c|c|c|c|c|c|}
\hline
TXS & $z_{\rm em}$ & $z_{\rm DLA}$ & N$_{\rm HI}$ & $\nu_{\rm 21cm}$ & Time & $\Delta V$ & $\tau_{\rm RMS}$ & $\int \tau d{\rm V}$ & $f$ & $\ts$ \\
name  &              &              & $\tnhi$~\cm  &   MHz            &  Hrs &   \kms\     & $\times 10^{-3}$ & \kms\                &     &  K \\
\hline
0620+389 & 3.46 & 2.031 & $2.0 \pm 0.5$ & 468.63 & 0.8 (GBT)~ & 3.1 & 6.7 & ~~~~$< 0.22$~~~~~ & 0.79 & ~~$> 400$~~~ \\
2039+187 & 3.05 & 2.192 & $5.0 \pm 1.0$ & 444.99 & 2.2 (GBT)~ & 1.6 & 5.7 & $0.597 \pm 0.053$ & 0.35 & $160 \pm 35$ \\
1048+347 & 2.52 & 2.241 & $3.5 \pm 0.5$ & 438.26 & 2.5 (GBT)~ & 3.3 & 2.5 & ~~~~$< 0.05$~~~~~ & 0.69 & ~~$> 2155$~~ \\
0311+430 & 2.87 & 2.289 & $2.0 \pm 0.5$ & 431.87 & 2.0 (GBT)~ & 1.7 & 1.7 & $0.818 \pm 0.085$ & 0.52 & ~$72 \pm 18$ \\
0229+230 & 3.42 & 2.620 & $2.0 \pm 0.5$ & 392.38 & 3.3 (GBT)~ & 7.5 & 7.8 & ~~~~$< 0.25$~~~~~ & 0.30 & ~~$> 140$~~~ \\
0229+230 & 3.42 & 2.683 & $5.0 \pm 1.0$ & 385.67 & 3.3 (GBT)~ & 7.6 & 8.8 & ~~~~$< 0.30$~~~~~ & 0.30 & ~~$> 270$~~~ \\
2039+187 & 3.05 & 2.752 & $5.0 \pm 1.0$ & 378.58 & 12  (WSRT) & 1.9 & 5.4 & ~~~~$< 0.12$~~~~~ & 0.35 & ~~$> 790$~~~ \\
1239+376 & 3.81 & 3.411 & $2.5 \pm 0.5$ & 322.01 & 22  (GMRT) & 7.6 & 0.8 & ~~~~$< 0.034$~~~~ & 0.54 & ~~$> 2330$~~ \\
\hline
\end{tabular}
\end{center}
Notes: The spin temperatures of the last column assume the DLA covering factors listed in the penultimate 
column; see Table~\ref{table:vlba} and the main text for discussion of the DLAs towards the extended 
sources TXS0229+230, TXS0311+430 and TXS2039+187. 
\end{table*}

\section{Data analysis and results}
\label{sec:data}

All GBT data were analysed using the {\sc AIPS++} single-dish package {\sc DISH}. After 
initial editing to excise RFI and Spectral Processor failures, the spectra were 
calibrated and averaged together to measure the quasar flux density. For each 10-second 
spectrum, a second-order spectral baseline was then fit to line- and RFI-free channels and 
subtracted out, during the process of calibration. The residual spectra were then averaged 
together with weights determined by the measured system temperatures. In a few cases, a 
first- or second-order polynomial was fit to, and subtracted from, this average spectrum to 
obtain the final \hii\ spectrum.

\begin{figure}
\centering
\epsfig{file=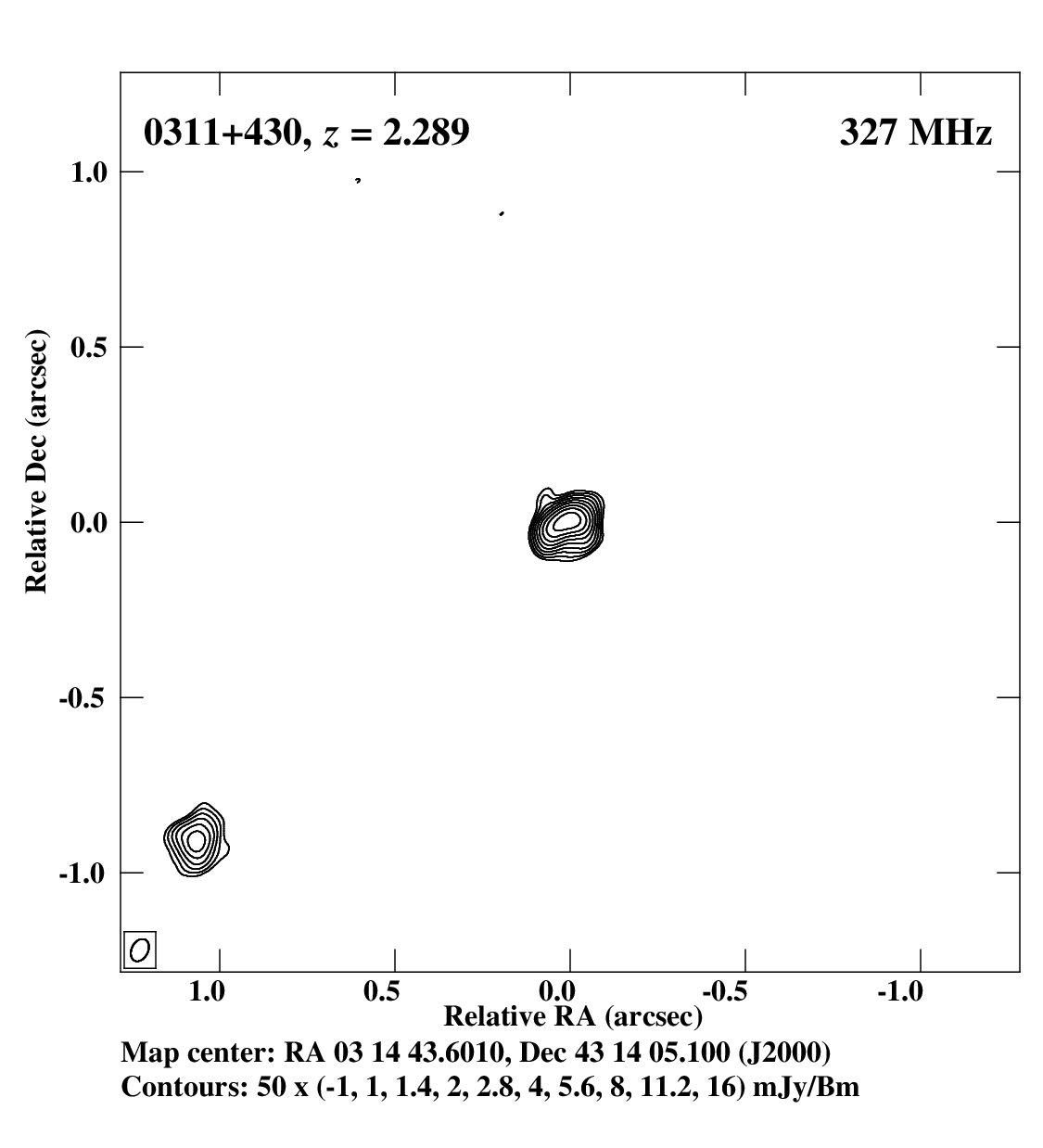,height=3.5truein,width=3.5truein}
\caption{Wide-field 327~MHz VLBA image of TXS0311+430, showing the radio component 
$1.42''$ to the southwest of the quasar core.}
\label{fig:0311vlba}
\end{figure}

The GMRT, WSRT and VLBA data were analysed in ``classic'' {\sc AIPS}. Data rendered unusable 
by various issues (e.g. dead antennas, correlator problems, RFI, etc) were first edited out, 
and standard calibration procedures then used to obtain the antenna-based complex gains. 
For each source, a number of channels were averaged to produce a ``channel-0'' dataset, 
and a series of self-calibration and imaging procedures used to iteratively determine the 
antenna gains and the final image, via standard procedures. 3-D imaging techniques were used 
for the GMRT data, sub-dividing the field into 37~facets to correct for the non-coplanarity 
of the array. For the GMRT and WSRT data, the continuum image was next subtracted out from 
the calibrated visibility data, after which a first-order polynomial was fit to the 
visibility spectra on each interferometer baseline and subtracted out. The residual U-V 
data were then shifted to the barycentric frame and imaged in all channels to obtain the 
final spectral cube. A cut through this cube at the quasar location yielded the \hii\ 
spectrum, after subtracting out a second-order baseline. 
The calibration steps for the VLBA data included ionospheric corrections and fringe-fitting 
for the delay rates, as well as the use of online measurements of the antenna gains and 
system temperatures to calibrate the flux density scale. The smaller number of VLBA antennas 
and the relatively poor ionospheric stability (due to which only 5-6 antennas could typically 
be retained) meant that the final VLBA images were produced with phase self-calibration 
alone. The task {\sc JMFIT} was used to measure the source flux density from these images, 
fitting an elliptical gaussian model to the radio core emission. For three sources, the entire
VLBA emission was found to be compact, with a single gaussian sufficient to model the emission. 
Extended emission was detected in the remaining three sources, TXS0229+230, TXS2039+187 
and TXS0311+430, and a 2-gaussian model was hence used here. In the case of the first two,
the extended radio structure is adjacent to the core emission; it is hence possible
that the foreground DLAs also cover the extended emission. Conversely, in the case of 
TXS0311+430, there is a second radio component located $1.42''$ to the southwest of the core 
emission, i.e. $\approx 11.7$~kpc at the DLA redshift; the foreground DLA is unlikely to 
cover this component unless the absorption arises in a large disk galaxy. We also note, 
in passing, that the core of TXS0311+430 also shows evidence for weak extended emission.
However, the poor U-V coverage implies that such extended features may arise close to bright 
continuum sources due to deconvolution errors. We have hence fit a single gaussian component
to the core emission of TXS0311+430, with a second gaussian fitted to the south-western 
radio component. The VLBA images of the six background quasar cores are displayed in 
Fig.~\ref{fig:vlba}, in order of increasing DLA redshift; further, a wide-field 
image of TXS0311+430, showing the southwest component, is shown in Fig.~\ref{fig:0311vlba}.
Note that, for the three sources with a 2-component Gaussian model, the more compact 
source component was identified with the radio core. 

The VLBA results are summarized in Table~\ref{table:vlba}; the columns of this table are:
(1)~the quasar name, (2)~the DLA redshift, (3)~the redshifted \hii\ line frequency, 
(4)~the VLBA observing frequency (MHz), (5)~the synthesized beam (in mas~$\times$~mas), 
(6)~the root-mean-square (RMS) noise (in mJy~beam$^{-1}$), (7)~the integrated flux 
density ($S_{\rm VLBA}$, in Jy) of the Gaussian components fit to the VLBA image, obtained 
using {\sc JMFIT}, (8)~the total source flux density $S_{\rm int}$ at the VLBA frequency, 
obtained either from direct measurements or by extrapolating from measurements at other 
frequencies [e.g. at 365~MHz \citep{douglas96} and 1400~MHz \citep{condon98}], (9)~the 
deconvolved angular size of the fitted components (in mas~$\times$~mas), (10)~the spatial 
extent of the core emission at the absorber redshift (in pc~$\times$~pc), and (11)~the 
covering factor $f$ of the foreground DLA, obtained by taking the ratio of the core flux 
density to the total source flux density $S_{\rm tot}$.

We note that it is difficult to estimate errors on the covering factor as the total flux 
density and VLBA measurements were not carried out simultaneously. Further, it should be 
emphasized that the deconvolved angular sizes (and hence, the estimates of the spatial 
extent of the core radio emission) are upper limits, as any residual phase errors would 
contribute to increasing the measured angular size.  The spatial extent of the radio core 
at the absorber redshift is $\la 1$~kpc in all cases, significantly smaller than the size 
of a galaxy.

Our estimates of the covering factor assume that the absorber only covers the 
compact VLBA core and does not cover any part of the extended radio emission. In the 
case of TXS0311+430, TXS2039+187 and TXS0229+230, which have extended structure in the 
VLBA images (on angular scales of $\sim 0.1-1''$, or $\sim 0.8-8$~kpc at the DLA redshift), 
it is possible that the foreground absorber may cover some of the extended emission, 
implying a higher covering factor. Our covering factor estimates for these DLAs should hence 
be treated as lower limits.

\begin{figure}
\centering
\epsfig{file=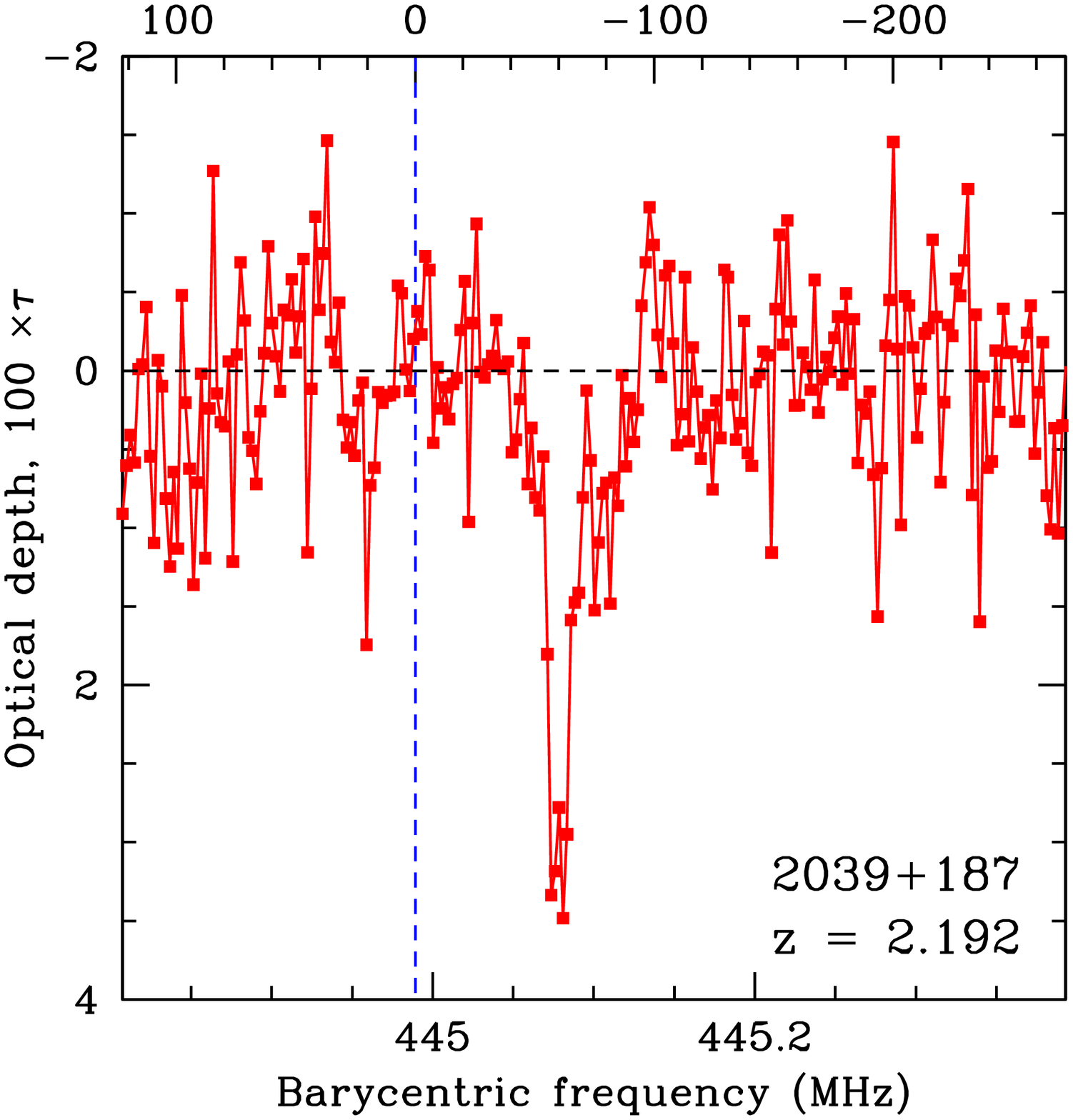,height=3.3truein,width=3.3truein}
\caption{A new detection of \hii\ absorption, at $z = 2.192$ towards TXS2039+187,
with \hii\ optical depth plotted against barycentric frequency, in MHz. The top 
axis shows velocity (in \kms), relative to the DLA redshift, while the \hii\ line 
frequency expected from the Ly$\alpha$ redshift is indicated by the vertical dashed line.
Note that the offset between the peak \hii\ and Ly-$\alpha$ absorption redshifts 
($\approx 60$~\kms) is smaller than the error in the Ly-$\alpha$ redshift ($\approx 180$~\kms).}
\label{fig:detect}
\end{figure}

\begin{figure*}
\centering
\epsfig{file=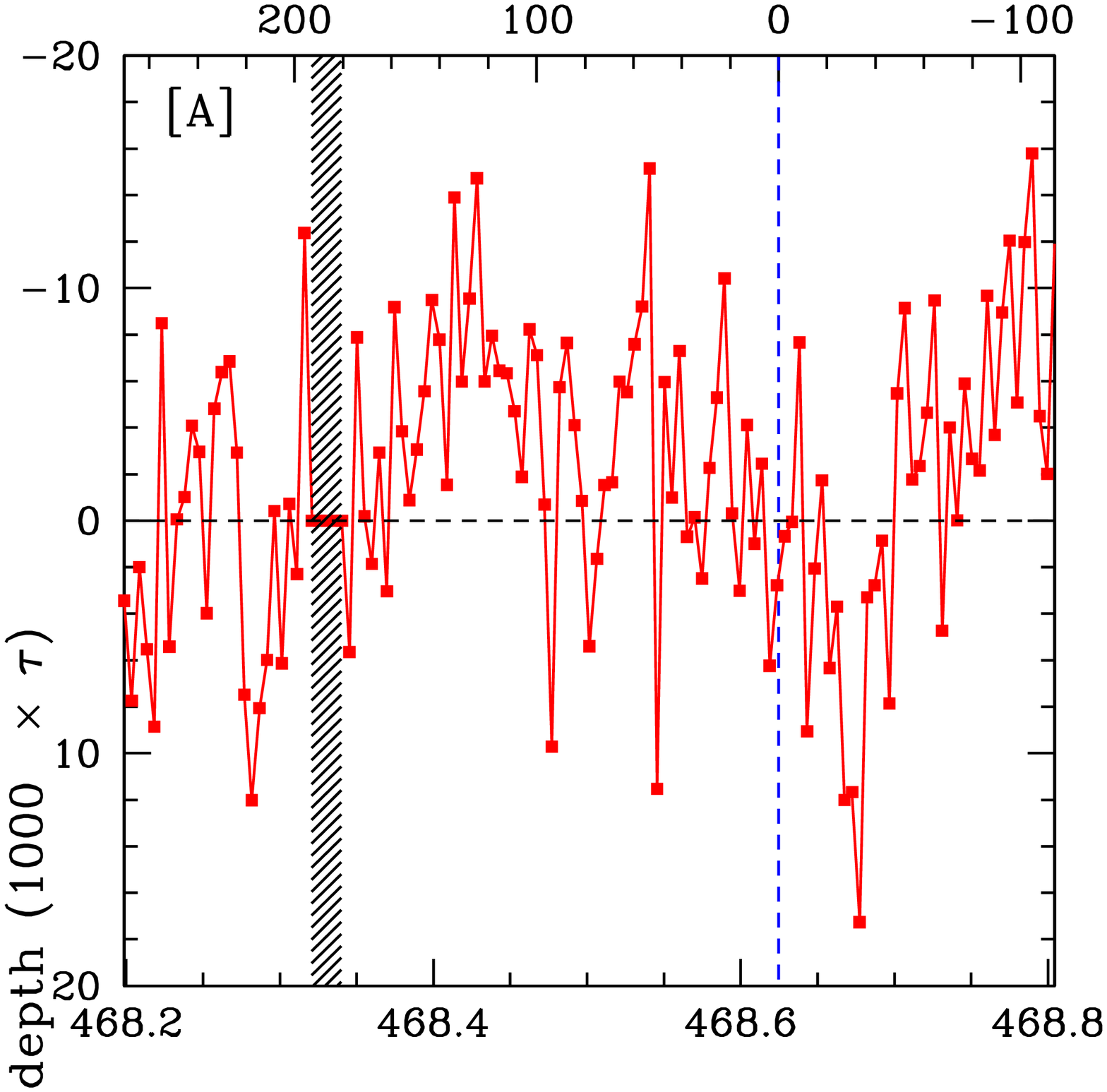,height=2.3truein,width=2.3truein}
\epsfig{file=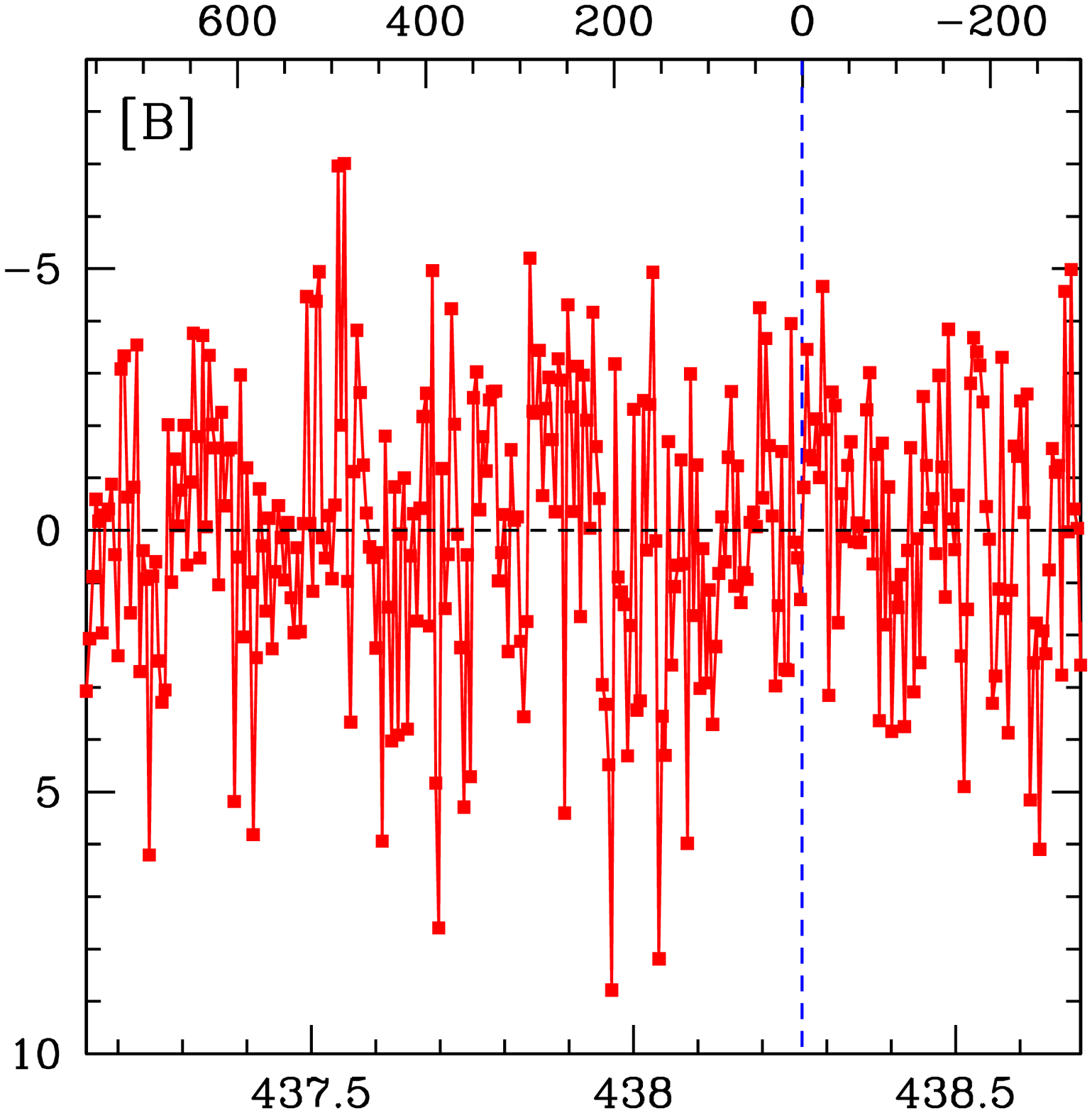,height=2.3truein,width=2.3truein}
\epsfig{file=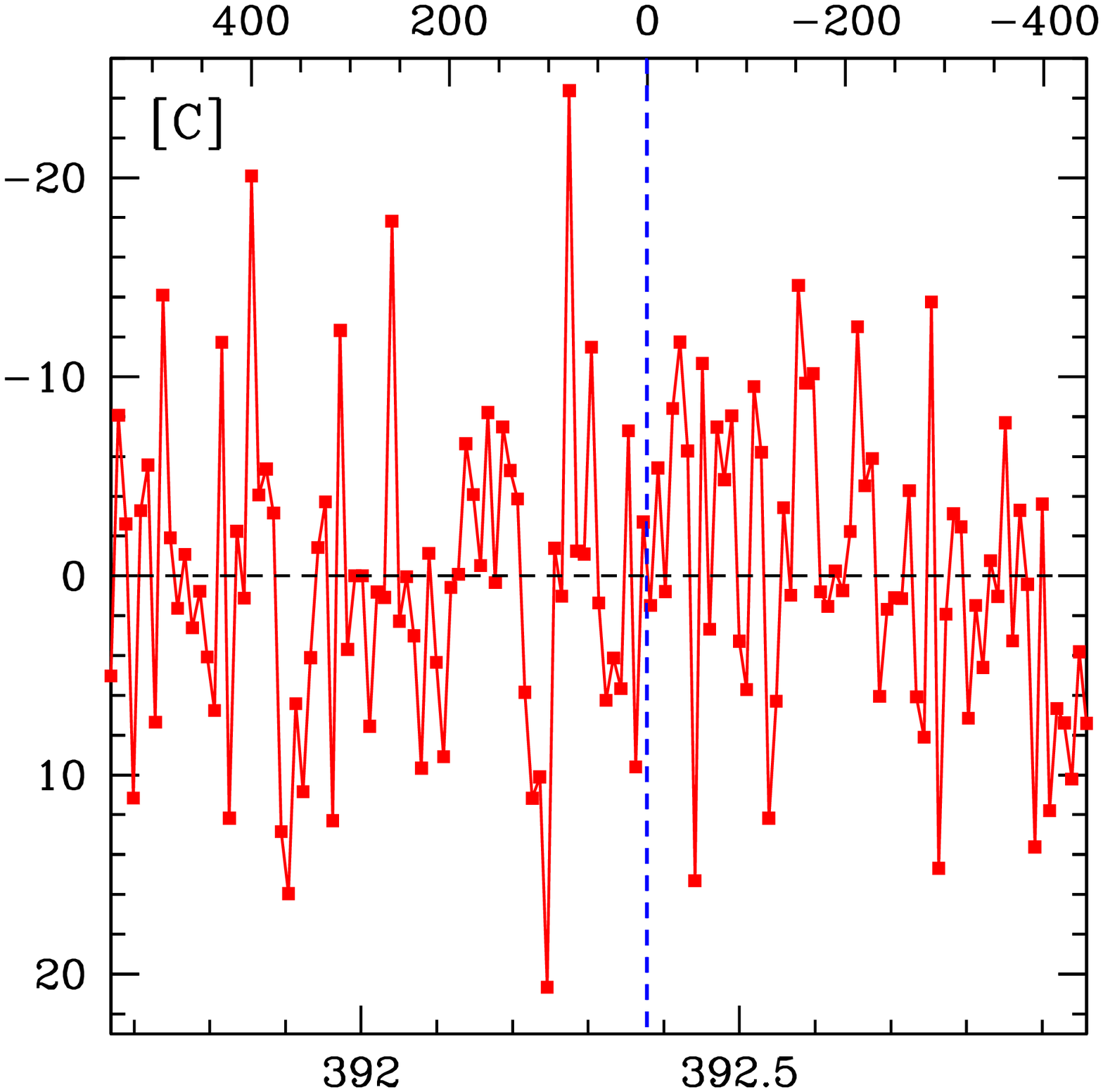,height=2.3truein,width=2.3truein}
\epsfig{file=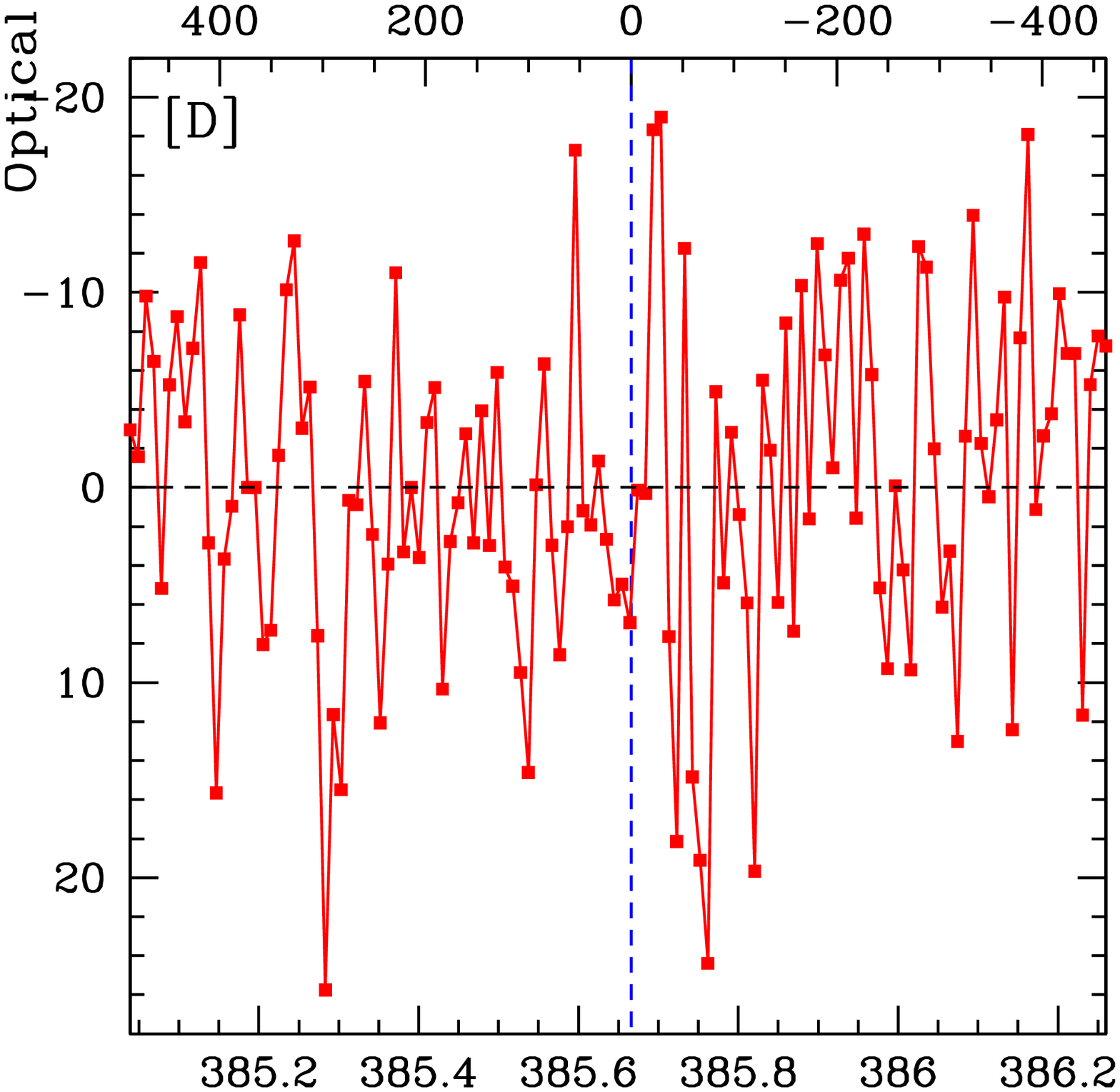,height=2.3truein,width=2.3truein}
\epsfig{file=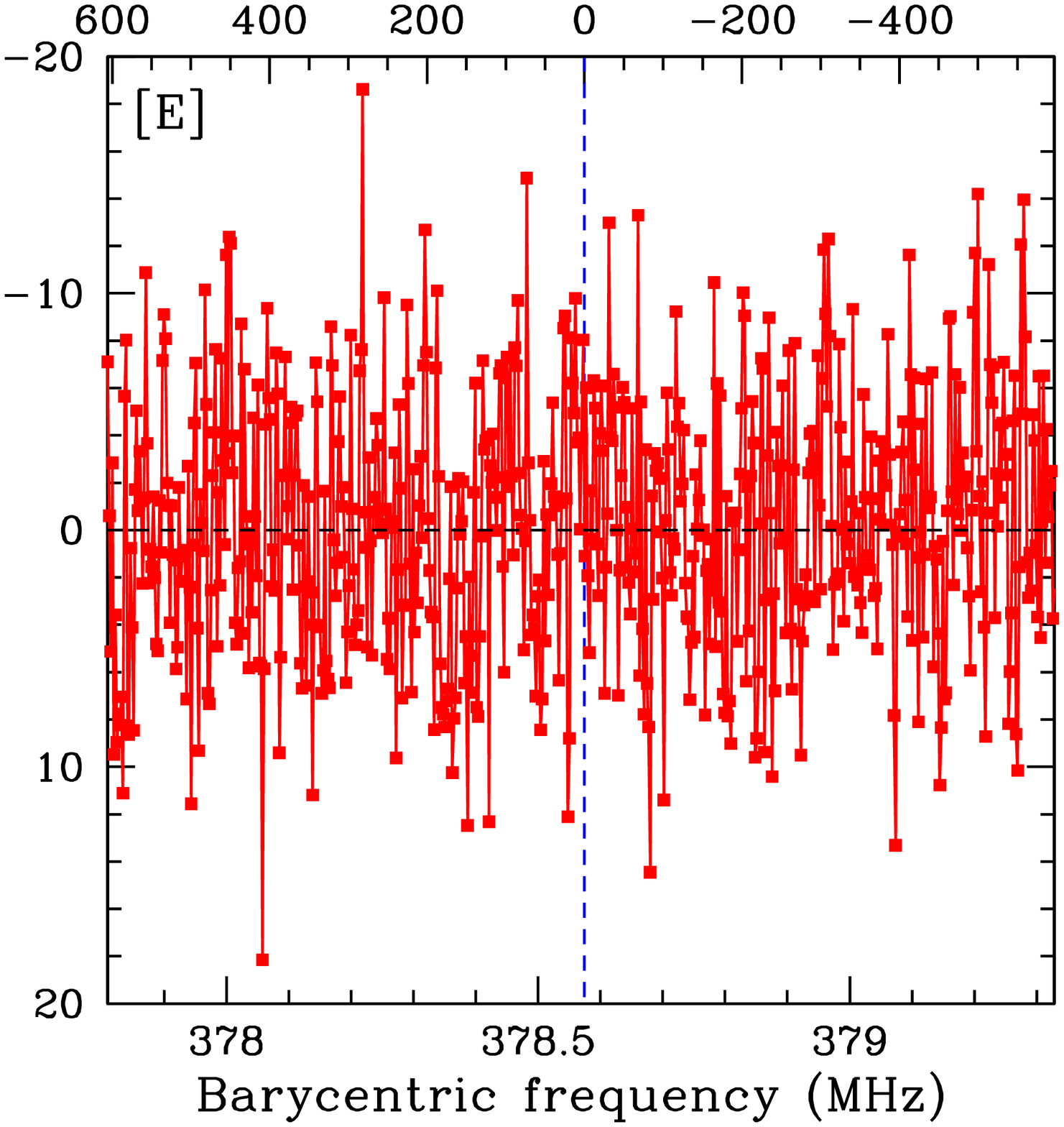,height=2.3truein,width=2.3truein}
\epsfig{file=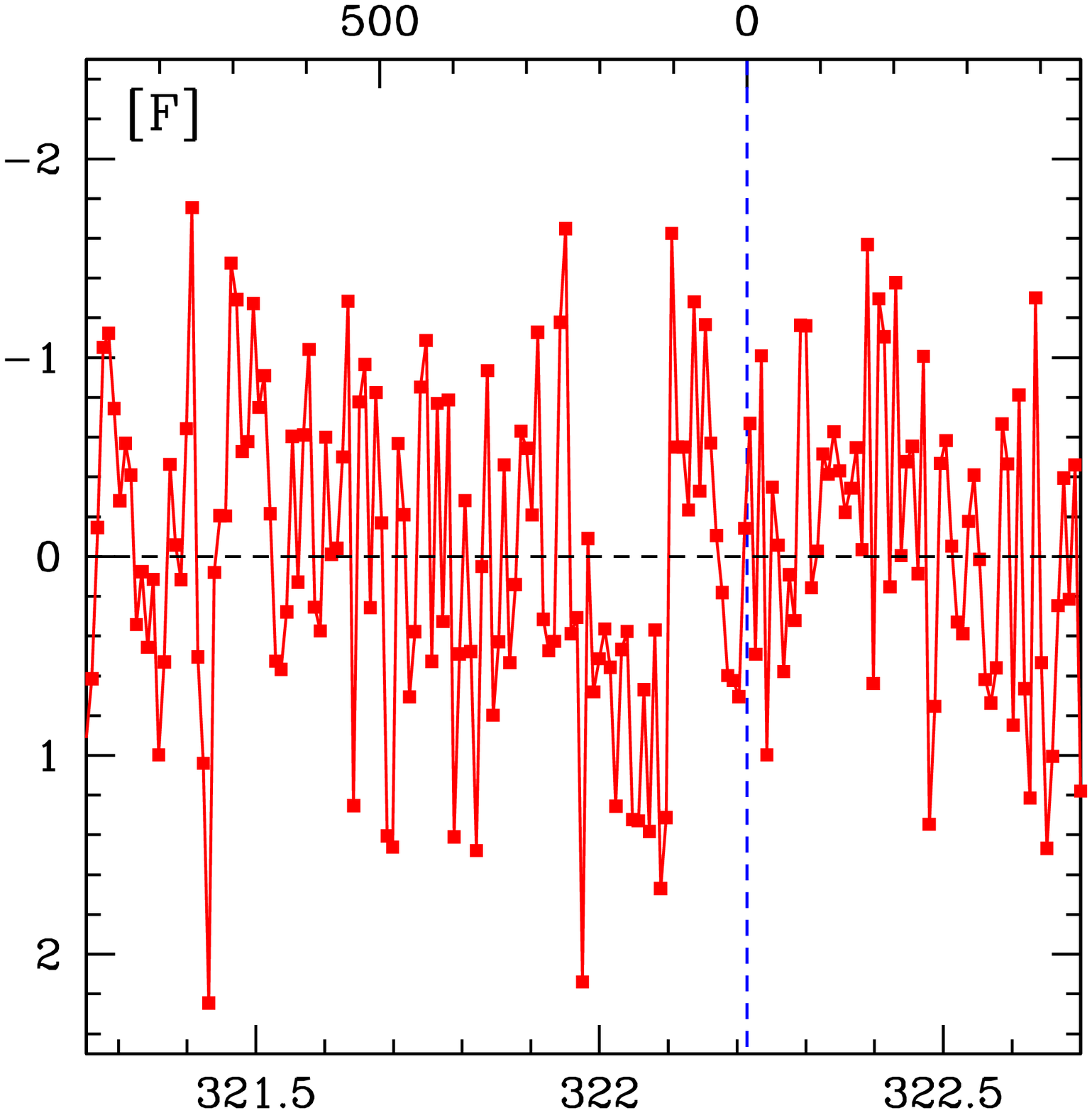,height=2.3truein,width=2.3truein}
\caption{The six non-detections of \hii\ absorption, with optical depth ($10^3 \times \tau$)
plotted against barycentric frequency, in MHz. The top axis in each panel shows velocity (in \kms), 
relative to the DLA redshift, while the expected \hii\ line frequency is indicated by the 
dashed line in each panel. The panels contain spectra for [A]~the $z=2.031$ DLA towards TXS0620+389 
(where the shaded region indicates a frequency range affected by RFI), [B]~the $z=2.241$ DLA 
towards TXS1048+347, [C]~the $z = 2.620$ DLA towards TXS0229+230, [D]~the $z=2.683$ DLA towards 
TXS0229+230, [E]~the $z=2.752$ DLA towards TXS2039+187, and [F]~the $z = 3.4082$ DLA towards TXS1239+376.
}
\label{fig:nondetect}
\end{figure*}

The DLAs at $z = 2.192$ towards TXS2039+187 and $z = 2.289$ towards TXS0311+430 showed 
strong \hii\ absorption features on multiple observing runs; the latter detection was 
reported, and discussed in detail, by \citet{york07}, and will not be described here. 
The \hii\ spectrum from the $z = 2.192$ DLA towards TXS2039+187 is shown in 
Fig.~\ref{fig:detect}; this has an integrated \hii\ optical depth of $0.597 \pm 0.053$~\kms, 
and a velocity spread (between nulls) of 55~\kms. The peak \hii\ absorption redshift in
this system ($z \approx 2.191$) lies between the metal-line and Ly-$\alpha$ absorption redshifts 
($z_{\rm HI} \approx 2.192$ and $z_{\rm metals} \approx 2.190$), although the errors on the 
latter are large due to the low resolution of the present optical spectrum. It would be interesting
to test whether there is indeed a velocity offset between the \hii\ and metal absorption, through a 
higher-resolution optical spectrum.

The spectra of the remaining six DLAs are shown in Fig.~\ref{fig:nondetect}, in order of increasing 
redshift. While weak absorption features are visible near the expected redshifted \hii\ 
line frequency in the spectra of TXS0620+389 and TXS1239+376, these are not statistically 
significant ($< 3\sigma$ and $\approx 3.5\sigma$, respectively), on smoothing to a resolution 
comparable to the width of the feature. A Kolmogorov-Smirnov rank-1 test finds the spectra 
consistent with a normal distribution, as would be expected from random noise. However, if one fits a 
linear baseline to the spectrum of the $z = 2.031$ DLA towards TXS0620+389 (after excluding the 
putative feature), and subtracts it out, the statistical significance of the feature increases to 
$\approx 5\sigma$. Unfortunately, there is very little spectral baseline at frequencies above 468.7~MHz 
and it is hence not possible to rule out the possibility that the apparent linear baseline arises 
from a ripple in the spectrum. We have hence conservatively chosen to quote a $3\sigma$ upper limit 
on the \hii\ optical depth of the DLA, based on the spectrum in Fig.~\ref{fig:nondetect}[A]. 
The other four DLAs showed no discernible absorption at or near the redshifted \hii\ line frequency. 

\citet{srianand12} have previously reported a GMRT non-detection of \hii\ absorption from 
the $z = 3.411$ DLA towards TXS1239+376. However, our longer integration time with the 
GMRT has allowed us to reach a deeper sensitivity. We also present an updated covering factor 
of $f = 0.54$ from our 327~MHz VLBA image, while \cite{srianand12} previously estimated $f = 1$ 
from a 1.4~GHz VLBA image. The difference is likely to be because compact radio cores typically 
have inverted spectra due to synchrotron self-absorption, while extended radio structures have steep
spectra and hence contribute a larger fraction of the emission at  lower frequencies. This 
emphasizes the need for VLBI estimates of the covering factor at frequencies close to
the redshifted \hii\ line frequency. Consequently, we caution that our 1.4~GHz VLBA 
estimate of the covering factor of the $z = 2.031$ DLA towards TXS0620+389 may be an 
over-estimate. All other DLAs of our sample have their covering factors measured at 
327~MHz, within a factor of 1.5 of the redshifted \hii\ line frequency.

Our final results are summarized in Table~\ref{table:sample} whose columns contain: 
(1)~the quasar name, (2)~the quasar redshift, (3)~the DLA redshift, (4)~the \hi\ column 
density, $\nhi$, from the Lyman-$\alpha$ absorption line \citep{ellison08}, (5)~the redshifted \hii\ 
line frequency, $\nu_{\rm 21cm}$ in MHz, (6)~the on-source time, in hours, with the telescope 
in parenthesis, (7)~the velocity resolution $\Delta V$, in \kms, (8)~the root-mean-square 
noise (RMS) at this resolution, in optical depth units, (9)~the integrated \hii\ optical depth 
$\int \tau d{\rm V}$, or, for non-detections, the $3\sigma$ upper limit on $\int \tau d{\rm V}$, 
in \kms, (10)~the covering factor $f$, and (11)~the spin temperature $\ts$, or, for non-detections, 
the $3\sigma$ lower limit on $\ts$, in K, computed from the expression 
$\nhi = 1.823 \times 10^{18} \left[ \ts/f \right] \int \tau d{\rm V}$. For non-detections, the line 
profile was assumed to be a Gaussian with full-width-at-half-maximum (FWHM)=15~\kms, similar to the 
widths of individual spectral components in redshifted \hii\ absorbers, with the RMS noise computed at 
a velocity resolution of $\approx 15$~\kms. This yields a conservative upper limit to the integrated \hii\ 
optical depth; note that the line FWHM for cold gas (kinetic temperature~$< 200$~K) due to thermal 
broadening is $\le 3$~\kms.

For six DLAs of the sample, our non-detections of \hii\ absorption imply that we only obtain 
lower limits to the spin temperature. Three of these have weak lower limits on the spin 
temperature, $\ts > (140 - 400)$~K, while the remaining three have high lower limits,
$\ts > 790$~K, similar to values obtained so far in the bulk of the high-$z$ DLA population 
(e.g. \citealt{kanekar03,kanekar09c}). However, the two DLAs of the sample with detections of 
\hii\ absorption, at $z = 2.192$ towards TXS2039+187 and $z = 2.289$ towards TXS0311+430, both 
have low spin temperatures, $\ts = (160 \pm 35) \times (f/0.35)$~K and $\ts = (72 \pm 18) \times 
(f/0.52)$~K, respectively. These are the first two DLAs with low spin temperatures ($\ts < 300$~K) 
at $z > 1$. Indeed, the spin temperature of the $z = 2.289$ DLA towards TXS0311+430 is the 
lowest obtained to date for any DLA, independent of redshift!

We note that the spin temperatures for the DLAs towards TXS2039+187, TXS0229+230 and TXS0311+430 
assume that the DLAs cover only the radio core, i.e. covering factors of $f=0.35$ (TXS2039+187), $f=0.30$ 
(TXS0229+230) and $f=0.52$ (TXS0311+430). If the foreground DLAs cover the entire flux density detected 
in the VLBA images, the spin temperature limits would be $\ts > 295$~K ($z = 2.620$ towards TXS0229+230), 
$\ts > 565$ ($z = 2.683$ DLA towards TXS0229+230) and $\ts > 1265$~K ($z = 2.752$ DLA towards TXS2039+187).
The spin temperature of the $z = 2.192$ DLA towards TXS2039+187 would be $\ts = (255 \pm 55) \times (f/0.56)$~K 
while that of the $z = 2.289$ DLA towards TXS0311+430 would be $\ts = (100 \pm 25) \times (f/0.72)$~K.

Finally, we cannot formally rule out the possibility that both VLBA components detected in the 
image of TXS0311+430 are actually compact steep-spectrum lobes and that the sightline towards 
the optical QSO lies in between the two lobes. If so, the spin temperature estimate for this sightline
could be incorrect, as the \hi\ column density determined towards the optical QSO need not be 
the same as that towards either of the radio lobes.

\section{Discussion}
\label{sec:ts}

The high derived spin temperatures in high-$z$ DLAs have been a puzzle ever since the first 
\hii\ absorption studies of DLAs at $z \gtrsim 2$ \citep{wolfe79,wolfe81}. Sightlines 
in the Milky Way and local spirals typically show low spin temperatures: more 
than 80\% of the sightlines through the Milky Way and M31 in the sample of \citet{braun92} 
have $\ts < 300$~K. Similarly, even high-latitude Milky Way sightlines with \hi\ column densities 
above the DLA threshold typically have $\ts \le 500$~K \citep{kanekar11b}. In the case of DLAs,
both high and low spin temperatures have been found in DLAs at $z \le 0.7$
\citep[e.g.][]{lebrun97,lane98,chengalur99,kanekar01a,ellison12}. In contrast, until the present sample, 
{\it every DLA at $z > 1$ had $\ts > 500$\,K, with the vast majority having $\ts \gtrsim 1000$\,K}
\citep{wolfe79,wolfe81,wolfe85,carilli96,briggs97,kanekar03,kanekar06,kanekar07,srianand10,srianand12}.

\citet{curran05} emphasized that the high $\ts$ estimates in high-$z$ DLAs could arise
because the low-frequency radio emission of the quasar is extended and hence not entirely 
covered by the foreground DLA. This is because most earlier studies of high-$z$ DLAs measured 
the ratio $\ts/f$, rather than $\ts$ itself. However, the low-frequency covering factors of 
a sample of high-$z$ DLAs were recently measured using very long baseline interferometric studies 
of the background radio quasars, and found to be $\ge 0.4$ in all cases \citep{kanekar09a}. 
Further, these authors found no difference between the distributions of the covering factors 
of the low-$z$ and high-$z$ DLA samples. This implies that low covering factors are not the 
cause of the high $\ts$ values obtained in high-$z$ DLAs. Note that the VLBA measurements 
of the covering factor of the DLAs of the present sample obtained $f \ge 0.3$ in all cases.

The simplest interpretation of the high spin temperatures obtained in high-$z$ DLAs is that 
the neutral hydrogen in the absorbers is mostly in the warm phase (the warm neutral medium; 
WNM), with very little cold \hi\ 
\citep{carilli96,chengalur00,kanekar03,kanekar09c}. A plausible explanation for this is a paucity of 
cooling avenues in the absorbers, due to their low metallicity \citep{kanekar01a}. In support 
of this hypothesis, a predicted anti-correlation between DLA metallicity and spin temperature has 
been recently discovered, with high $\ts$ values prevalent in low-metallicity DLAs and low 
$\ts$ values in high-metallicity systems \citep{kanekar09c,ellison12}. Note that \citet{carswell12} 
obtain a high WNM fraction in a low-metallicity DLA, at $z = 2.076$ towards QSO2206$-$199. 
Given the above anti-correlation, high-metallicity DLAs would be expected to have sizeable 
fractions of cold \hi, and hence low spin temperatures, even at high redshifts. However, although 
about 10\% of DLAs at $z > 2$ have high metallicities ([Z/H]~$\ge -0.5$; e.g.  \citealt{prochaska07}), 
low spin temperatures have not hitherto been obtained at high redshifts. On the other hand, we note 
that high metallicity DLAs do indeed show a higher probability of detection of H$_2$ 
absorption \citep[e.g.][]{petitjean06,noterdaeme08}, consistent with the hypothesis that such 
DLAs have significant cold neutral medium (CNM) fractions.

The absorbers towards TXS2039+187 and TXS0311+430 are the first two DLAs at $z > 1$ with low 
spin temperatures, $\ts < 300$~K. While the detection of \hii\ absorption in the latter system was 
reported by \citet{york07}, no VLBI studies of the background quasar were available then.
\citet{york07} hence used a GMRT 602~MHz image to estimate a covering factor of $\approx 1$. Our 
VLBA observations yield the lower limit $f > 0.52$ on the covering factor, assuming that the core 
alone is covered by the foreground absorber, and confirm the low estimated spin temperature 
in this DLA. We note that the spin temperature estimate in this DLA is lower by a factor of 
$\approx 2$ than the original estimate of \citet{york07}, emphasizing the need for VLBA imaging 
studies at frequencies close to the redshifted \hii\ line frequency. 

The low $\ts$ values in the DLAs towards TXS2039+187 and TXS0311+430 indicate that the two absorbers 
are likely to contain significant fractions of the cold neutral medium, $\gtrsim 50$\% of the total 
\hi\ content (as $\ts$ is the column-density weighted harmonic mean of spin temperatures in different 
phases along the sightline; e.g. \citealt{kanekar04}). The anti-correlation between spin temperature 
and metallicity \citep{kanekar09c,ellison12} implies that both these DLAs should have high metallicities, 
[Z/H]~$\ge -0.5$; conversely, the DLAs with high spin temperatures would be expected to have 
low metallicities, [Z/H]~$< -1$. As expected, the absorber towards TXS0311+430 has [Si/H]~$> -0.6$, among 
the highest metallicities of DLAs at $z > 2$ \citep{york07,ellison08}, while, in the case of the 
system towards TXS2039+187, \citet{ellison08} obtain [Si/H]~$> -1.4$. As noted by \citet{ellison08}, 
both of these estimates are lower limits as they are based on low-resolution ($\approx 5\AA$) spectra 
and assume that the Si{\sc ii}-$\lambda$1808 absorption lines are unsaturated. It would be interesting 
to accurately determine the metallicity of the DLA towards TXS2039+187 from a high-resolution spectrum.

If the two DLAs do indeed have high metallicities, the correlation between velocity spread and 
metallicity found in high-$z$ DLAs \citep{ledoux06,prochaska08}, interpreted as arising 
due to an underlying mass-metallicity relation, suggests that the two absorbers are likely to be 
massive galaxies. Note that the velocity spread (between nulls) of the \hii\ absorption in TXS0311+430 
is indeed quite large, $130$~\kms, as would be expected of a large galaxy; it is thus plausible that 
the absorber also covers the south-west VLBA component. If so, it might be possible to map the \hii\ absorption
against the VLBA structure and directly estimate the transverse extent of the DLA \citep[e.g.][]{kanekar04,kanekar05b}. 
Conversely, the \hii\ absorption towards TXS2039+187 has a velocity spread of only $55$~\kms; if this 
is a massive galaxy, it is likely to be a relatively face-on system. Follow-up imaging studies of 
the host galaxies of the two DLAs will be of much interest. 
  

In summary, we report a deep search for redshifted \hii\ absorption in eight DLAs at $z > 2$, 
with the GBT, the GMRT and the WSRT, allied with low-frequency VLBA imaging of the quasar radio 
emission to determine the DLA covering factors. High spin temperature limits, $\ts \ge 790$~K, 
were obtained in three absorbers, and weak lower limits, $\ts > 140-400$~K, in three systems.
We report a new detection of \hii\ absorption in the $z = 2.192$ DLA towards TXS2039+187, 
only the sixth case of \hii\ absorption in DLAs at $z > 2$. The detection of \hii\ absorption 
in the eighth DLA, at $z= 2.289$ towards TXS0311+430, was earlier reported by \citet{york07},
but our new low-frequency covering factor measurement allows a better estimate of the DLA 
spin temperature. These are the first two DLAs at $z > 1$ with low spin temperatures, 
$\ts < 300$~K. The fraction of DLAs with low spin temperatures ($\approx 7$\%) at $z > 2$ is 
similar to that of DLAs with high metallicity, [Z/H]~$\ge -0.5$ ($\approx 8$\%). This supports 
the hypothesis that the high spin temperatures of most high-$z$ DLAs arise because the low 
absorber metallicities imply fewer routes for gas cooling.

\section*{Acknowledgements}

We thank the staff of the GMRT, the GBT, the VLBA and the WSRT, who have made these 
observations possible. The GMRT is run by the National Centre for Radio Astrophysics 
of the Tata Institute of Fundamental Research. The WSRT is operated by ASTRON (the 
Netherlands Institute for Radio Astronomy), with support from the Netherlands 
Foundation for Scientific Research (NWO). The NRAO is a facility of the National 
Science Foundation operated under cooperative agreement by Associated Universities, 
Inc.. NK acknowledges support from the Department of Science and Technology through 
a Ramanujan Fellowship. We thank an anonymous referee for comments and suggestions 
that have improved this manuscript.

\bibliographystyle{mn2e}
\bibliography{ms}

\label{lastpage}

\end{document}